\documentclass[english,keywords,amsmath,amssymb,twocolumn]{revtex4}
\usepackage[T1]{fontenc}
\usepackage[latin1]{inputenc}
\usepackage{braket}
\usepackage{babel}%
\usepackage{graphicx}
\usepackage{color}
\usepackage{bm}
\usepackage{longtable}
\usepackage{amsmath}
\usepackage{amsfonts}
\usepackage{dsfont}
\usepackage{amssymb}
\usepackage{hyperref}
\usepackage{centernot}

\begin{document}

\title{ Two definitions of maximally  $\psi$-epistemic ontological model and preparation non-contextuality }
\author{A. K. Pan \footnote{akp@nitp.ac.in}}
\affiliation{National Institute of Technology Patna, Ashok Rajhpath, Patna 800005, India}
\begin{abstract}
The notion of $\psi$-epistemicity in an ontological model of quantum theory and its relation to non-contextuality has recently been studied. An ontological model is termed as maximally $\psi$-epistemic if  the overlap between any two quantum states is fully accounted for by the overlap of their respective probability distributions of ontic states. However, in literature, there exists the two different mathematical definitions (termed here as 1M$\psi$E and 2M$\psi$E) that capture the equivalent notion of maximal $\psi$-epistemicity. In this work, we provide three theorems to critically examine the connections between preparation non-contextuality and the aforementioned two definitions of maximal $\psi$- epistemicity.   In Theorem-1, we provide a simple and direct argument of an existing proof to demonstrate that the mixed state preparation non-contextuality implies the first definition of maximal $\psi$-epistemicity.  In Theorem-2, we prove that the second definition of maximal $\psi$-epistemicity implies pure-state preparation non-contextuality. If both the definitions capture the equivalent notion of maximal $\psi$-epistemicity then from the aforementioned two theorems one infers that the mixed-state preparation non-contextuality implies pure-state preparation non-contextuality. But, in Theorem-3, we demonstrate that the mixed-state preparation non-contextuality in an ontological model implies pure-state contextuality and vice-versa. This leads one to conclude that 1M$\psi$E and 2M$\psi$E capture inequivalent notion of maximal $\psi$-epistemicity. We further discuss the implications of the three theorems in an ontological model.  
\end{abstract}
\maketitle

Since the inception of quantum theory it remains a debatable issue among physicists  and philosophers whether a quantum state directly corresponds to an element of reality or  an abstract entity provides subjective knowledge about some underlying physical reality. An elegant mathematical and conceptual framework of ontological models of an operational theory have recently been introduced by Harrigan and Spekkens\cite{harrigan} which provided a precise approach to examine those questions from modern perspective. Two major distinctions between ontological models of quantum theory have been made, referred to as $\psi$-ontic and $\psi$-epistemic views of quantum state. While the former interprets the quantum state as a part of reality, the latter is in favor of interpreting the quantum state as merely the knowledge of the reality but not the reality itself. 

Before proceeding further, let us encapsulate the notion of ontological model of an operational theory, coherently put forwarded in  \cite{harrigan,spek05} and  adopted by many authors. Given a preparation procedure $P\in \mathcal{P}$ and a measurement procedures $M\in \mathcal{M}$, an operational theory assigns probability $p(k|P, M)$ in obtaining a particular outcome $k\in \mathcal{K}_{M}$. Here, $\mathcal{M}$ and $\mathcal{P}$ are the set of measurement and preparation procedures respectively. In quantum theory, a preparation procedure $(P)$ produces a density matrix $\rho$ and measurement procedure $(M)$ in general described by a suitable set of POVM elements $\{E_k\}$. The probability of occurrence of a particular outcome $k$ is obtained via the Born rule, i.e., $p(k|P, M)=Tr[\rho E_{k}]$. 

Ontological model of quantum theory assumes that whenever $\rho$ is prepared by a preparation procedure $P\in \mathcal{P}$, a probability distribution (termed as epistemic state) $\mu_{P}(\lambda|\rho)$ of ontic states $\lambda$ is prepared, where $\lambda \in \Lambda$ and $\Lambda$ is the ontic state space. Given a measurement operator $E_{k}$, realized through a measurement procedure $M \in \mathcal{M} $, the ontic state  $\lambda$ assigns a response function $\xi_{M}(k|E_{k}, \lambda) \in [0,1]$ satisfying $\sum_{k}\xi_{M}(k|\lambda, E_{k})=1$. A viable ontological model of quantum theory has to reproduce the Born rule, i.e., $\forall \rho $, $\forall E_{k}$ and $\forall k$, $\int _\Lambda \mu_{P}(\lambda|\rho) \xi_{M}(k|\lambda, E_{k}) d\lambda =Tr(\rho E_{k})$. For the case of pure state $\rho=|\psi\rangle\langle\psi|$ and projective measurement $E_{k}=\Pi_{\phi}=|\phi\rangle\langle\phi|$, we have $\int _\Lambda \mu(\lambda|\psi) \xi(\phi|\lambda, \Pi_{\phi}) d\lambda =|\langle\phi|\psi\rangle|^2$. Ontological models successfully reproduce quantum probabilities unless additional constraints ( such as, non-contextuality and locality) are imposed. Note here that if the epistemic state $\mu_{P}(\lambda|\rho)$ non-trivially depends on the preparation procedure $P$ then the ontological model is preparation contextual. Similarly for the response function. One may skip mentioning the dependence of response function on $P$  when the issue of contextuality is not relevant.

The traditional Kochen and Specker (KS) theorem \cite{kochen} demonstrates that a deterministic measurement non-contextual  model cannot reproduce all the predictions of quantum theory. However, the KS theorem has a limited scope of applicability. The notion of non-contextuality was generalized by Spekkens \cite{spek05} for any arbitrary operational theory and extended the formulation to the transformation and preparation procedures. Following the formulation in \cite{spek05}, an ontological model of an operational theory is assumed to be non-contextual as follows. If two experimental procedures are equivalent in an operational theory then they are represented equivalently in the relevant ontological model. 

More specifically, an ontological model of operational quantum theory is assumed to be measurement non-contextual if $\forall P: \ \  p(k|P, M)=p(k|P, M^{\prime})\Rightarrow \ \ \forall \lambda :\ \  \xi_{M}(k|\lambda, E_{k})=\xi_{M^{\prime}}(k|\lambda, E_{k})$, where $M$ and $M^{\prime}$ are two measurement procedures realizing the same POVM element $E_k$.  Note that, the traditional KS non-contextuality \cite{kochen} assumes the measurement non-contextuality for projectors along with their deterministic response functions. Along the same line,  an ontological model of quantum theory is assumed to be preparation non-contextual \cite{spek05} if  	$\forall M$ :
\begin{align}
p(k|P,M)=p(k|P^{\prime},M)\Rightarrow \mu_{P}(\lambda|\rho)=\mu_{P^{\prime}}(\lambda|\rho)
\end{align}
where $P$ and $P^{\prime}$ are two distinct procedures for preparing the same density matrix $\rho$. 

For better readability, some clarifications  regarding the applicability of preparation non-contextuality assumption will be helpful. Usually in the literature, the assumption of preparation non-contextuality ($\mu_{P}(\lambda|\rho)=\mu_{P^{\prime}}(\lambda|\rho)$) corresponds to a mixed state $\rho$ prepared by two or more distinct preparation procedures. However, the same assumption, although hitherto not explicitly used, should equally be applicable to a pure state $\rho=|\psi\rangle\langle\psi|$ prepared by two or more preparation procedures. %We note here that any logical all-versus-noting proof of preparation contextuality requires an inherent interplay between the preparation non-contextuality for mixed and pure states. In other words, there are scenarios when preparation non-contextuality cannot be assumed for both pure and mixed states.  To install the assumption of preparation non-contextuality for mixed state in that scenario the preparation contextuality for pure states needs to be assumed and vice-versa. 
For convenience, throughout this paper, we explicitly mention whether preparation non-contextuality  is assumed for a pure state or a mixed state.  

%In an ontological model of QM, the preparation non-contextuality for mixed state implies the outcome determinism for sharp measurements \cite{spek05}. 

The purpose of this work is to examine the connections between  preparation non-contextuality and $\psi$-onticity/$\psi$-epistemicity in an ontological model of quantum theory. A model is termed as $\psi$-ontic if an ontic state $\lambda$ can uniquely specifies the quantum state.  More specifically, for \emph{any} pair of different pure quantum states $|\psi_1\rangle$ and $|\psi_2\rangle$, if $\Lambda_{\psi_1}\cap\Lambda_{\psi_2}=\emptyset$, then the model is $\psi$-ontic \cite{harrigan}. This implies that different quantum states compatible with  disjoint set of ontic states. Here, $\Lambda_{\psi_1}$ and $\Lambda_{\psi_2}$ are the ontic state spaces corresponding to the quantum states $|\psi_1\rangle$ and $|\psi_2\rangle$ respectively.  If a model is not $\psi$-ontic then it is $\psi$-epistemic. In such a case at least a single ontic state is compatible with more than one quantum state. A model is thus said to be $\psi$-epistemic if there exists  \emph{at least } two quantum states $|\psi_1\rangle$ and $|\psi_2\rangle$ for which $\Lambda_{\psi_1}\cap\Lambda_{\psi_2}\neq\emptyset$. For a rigorous review of this topic, we refer Ref.\cite{leiferquanta} to the reader.  

The $\psi$-epistemic model is believed to provide reasonable explanation to many quantum phenomena including the measurement problem \cite{spek07}. But,  an interesting no-go theorem \cite{pbr} was recently proposed suggesting that the epistemic view of quantum state (satisfying an additional assumption of preparation independence of ontic states for composite systems) fails to reproduce quantum theory. Since then, several no-go theorems have been proposed using different forms of auxiliary assumptions \cite{sch12,aaronson,patra,hall,hardy}. Experimental verifications have also been reported \cite{nigg,ring}. Lewis \emph{et al.} \cite{lewis} have argued that if the aforementioned preparation independence is not assumed, $\psi$-epistemic model can reproduce quantum theory.   However, it is further proved \cite{barrett14, leifer14, bran} that $\psi$-epistemic ontological model cannot successfully  explain the distinguishability between two non-orthogonal quantum states belongs to the Hilbert space having dimension $d\geq 3$. 

By considering the degree of epistemicity, a stronger notion of epistemic view of quantum state was introduced \cite{maroney12} known as maximal $\psi$-epistemicity.  %Conceptually, in such a model the indistinguishability of \emph{any} two quantum states is fully accounted for by the overlap between the corresponding epistemic states of the ontic states. 
Conceptually, in a maximally $\psi$-epistemic model  the overlap between any two quantum states is fully accounted for by the overlap of their respective epistemic states in an ontological model. Importantly, there exists two different mathematical definitions of conceptually equivalent notion of maximal $\psi$-epistemicity.  As defined in \cite{maroney12}, for any two quantum states $|\phi\rangle $ and $|\psi\rangle$, the maximally $\psi$-epistemic model satisfies the condition $\int _{\Lambda_{\phi}} \mu(\lambda|\psi)  d\lambda = |\langle\phi|\psi\rangle|^2$  where $\Lambda_{\phi}$ is the support of $|\phi\rangle$ in ontic state space.  We call this as first definition of maximally $\psi$-epistemic model and henceforth denote as 1M$\psi$E .  By considering the quantum overlap between two arbitrary quantum states  and suitably defined classical overlap between the epistemic states in ontic state space corresponding to two quantum states, the second definition of maximally $\psi$-epistemic  model (henceforth, 2M$\psi$E)  is also introduced in \cite{maroney12} as a noise tolerant formulation of 1M$\psi$E and used in \cite{barrett14, leifer14, bran}. As we provide the explicit mathematical statements shortly, we note here that from both the mathematical definitions of maximally $\psi$-epistemicity,  it is proved \cite{maroney12, barrett14, leifer14, bran,ring}  that such a model cannot reproduce all quantum statistics if the Hilbert space dimension is grater than two. However, the critical examination regarding the equivalence (or inequivalence) between 1M$\psi$E and 2M$\psi$E has not been made.

A connection between the 1M$\psi$E and other no-go theorems, such as, KS theorem\cite{kochen} and preparation contextuality  \cite{spek05}  has been studied by Leifer and Maroney \cite{leifer13}. In particular, they showed that i) a preparation non-contextual model is 1M$\psi$E but converse does not hold good due to the existance of KS model in two dimension, and ii) a 1M$\psi$E model is KS non-contextual and again the implication is one directional.  It is crucial to note here that  assumption of preparation non-contextuality in their proof corresponds to  \emph{mixed state} in a qubit system \cite{leifer13}. Further the argument was generalized for the any mixed state belongs to arbitrary arbitrary dimensional system \cite{banik,leiferquanta}.

In this paper,  we provide three theorems regarding the relationships between preparation non-contextuality (for pure and mixed states) and the two definitions of maximal $\psi$-epistemicity. We show that putting those three theorems together leads a conceptual  inconsistency. In first theorem we provide a simple and elegant proof of \cite{leifer13} to show that mixed-state preparation non-contextuality implies 1M$\psi$E. Using the other definition of maximal $\psi$-epitemicity, we demonstrate the second theorem to prove that a 2M$\psi$E model is the pure-state preparation non-contextual one. In the third theorem we show that for imposing the assumption of  mixed-state  preparation non-contextuality one has to allow the \emph{preparation contextuality} for pure state.  Note that, if 1M$\psi$E and 2M$\psi$E capture the equivalent notion of maximal $\psi$-epistemicity, then first two theorems implies that the mixed-state preparation non-contextuality implies the pure-state preparation non-contextuality. This clearly contradicts  the statement of the third theorem.  One can then infer that the notion of maximal $\psi$-epistemicity captured by 1M$\psi$E and 2M$\psi$E models are conceptually inequivalent and 2M$\psi$E is not merely a noise tolerant version of 1M$\psi$E. 

\section{ mixed-state preparation non-contextuality $\Rightarrow$ 1M$\psi$E }

%It is interesting to note that, the Kochen-Specker model for $d=2$ admits M$\psi$E model. However,  M$\psi$E cannot be constructed for Hilbert space of dimensions three or more for any arbitrary pair of states due to the fact that for $d\geq3$ any deterministic ontological model is Kochen-Specker contextual\cite{leifer13}. However, there is no restriction for outcome indeterministic model. 
We start by encapsulating the first definition of M$\psi$E first introduced in \cite{maroney12}. For arbitrary two quantum states $|\phi\rangle $ and $|\psi\rangle$, by the definition of 1M$\psi$E model, the following condition holds.

\begin{align}
	\int _{\Lambda_{\phi}} \mu(\lambda|\psi)  d\lambda = |\langle\phi|\psi\rangle|^2 
\end{align}
where $\Lambda_{\phi}$ is the support of $|\phi\rangle$ in ontic state space. This model is outcome deterministic as $\xi(\phi|\lambda, \Pi_{\phi})=1$ for every $\lambda\in\Lambda_{\phi}$.  One can then write
\begin{align}
\label{born2}
\int _{\Lambda_{\phi}} \mu(\lambda|\psi) \xi(\phi|\lambda, \Pi_{\phi}) d\lambda =|\langle\phi|\psi\rangle|^2
\end{align}

Then, for a non-maximally or simply $\psi$-epistemic model,
\begin{align}
\label{born3}
	\int _{\Lambda_{\phi}} \mu(\lambda|\psi) \xi(\phi|\lambda, \Pi_{\phi}) d\lambda \leq \int _\Lambda \mu(\lambda|\psi) \xi(\phi|\lambda, \Pi_{\phi}) d\lambda
\end{align}
 Clearly, the model is 1M$\psi$E only if the equality in Eq.(\ref{born3}) holds. One can define a degree of epistemicity $f(\psi,\phi)$, so that, 
\begin{align}
\int _{\Lambda_{\phi}} \mu(\lambda|\psi) d\lambda = f(\psi,\phi) |\langle\psi|\phi\rangle|^{2}
\end{align}
where $0\leq f(\psi,\phi)\leq 1$. As the name suggests, the degree of epistemicity is maximum ( i.e., $f(\psi,\phi)=1$) in a 1M$\psi$E model where $|\psi\rangle$ and $|\phi\rangle$ are arbitrary and belong to a $d$-dimensional Hilbert space. Other conceptual features of such model is discussed in \cite{bal}. In an interesting work Leifer and Maroney \cite{leifer13} argued that a preparation non-contextual model is 1M$\psi$E but converse is not true due to the counterexample of KS model in two dimension. We note here again that the assumption of preparation non-contextuality  in \cite{leifer13} corresponds to a mixed-state. They proved the following theorem.\\
  
\textbf{Theorem-1:} \textit{If an ontological model of quantum theory is mixed-state preparation non-contextual then it is 1M$\psi$E.}\\

They started the argument by assuming an ontological model is non-maximally $\psi$-epistemic and proved that such a model is mixed-state preparation-contextual. We provide a direct and simpler proof of Theorem-1 shortly after recapitulating the essence of their proof. 

 Consider a maximally mixed qubit state $\rho=\mathcal{I}/2$ prepared in two different decompositions 

\begin{eqnarray}
\label{ms1}
\frac{\mathcal{I}}{2}&=&\frac{1}{2}\big(|\chi\rangle\langle \chi|+ \frac{1}{2}|\chi_{\perp}\rangle\langle \chi_{\perp}|\\
\label{ms2}
&=& \frac{1}{2}|\eta\rangle\langle \eta|+ \frac{1}{2}|\eta_{\perp}\rangle\langle \eta_{\perp}| 
\end{eqnarray}
This can be viewed as preparing $\rho=\mathcal{I}/2$  by using two distinct preparation procedures $P$ and $P^{\prime}$ respectively.  In an ontological model of quantum theory, the associated epistemic states   can be written as
\begin{eqnarray}
\label{ep1}
\mu_{P}(\lambda|\mathcal{I}/2)=(\mu_{\chi}+\mu_{\chi_{\perp}})/2\\
\label{ep2}
\mu_{P^{\prime}}(\lambda|\mathcal{I}/2)=(\mu_{\eta}+\mu_{\eta_{\perp}})/2
\end{eqnarray}
 and  $\Lambda_{P}=\lambda_{\chi}\cup\lambda_{\chi_{\perp}}$ and $\Lambda_{P^{\prime}}=\lambda_{\eta}\cup\lambda_{\eta_{\perp}}$  are the respective ontic state spaces. The assumption of mixed-state preparation non-contextuality implies $\Lambda_{P}=\Lambda_{P^{\prime}}$.

 Now, if the ontological model is \emph{not} 1M$\psi$E, then from Eq.(\ref{born3}) we can be write,  $\int _{\Lambda_{\eta}} \mu(\lambda|\chi) \xi(\eta|\lambda, \Pi_{\eta}) d\lambda < \int _\Lambda \mu(\lambda|\chi) \xi(\eta|\lambda, \Pi_{\eta}) d\lambda$. This means there is a set of ontic states $\Omega$, so that $\Lambda_{\chi}\cap \Omega=\emptyset$, and  $\xi(\eta|\lambda)>0$ for all $\lambda\in \Omega$. Since $|\chi\rangle\perp|\chi_{\perp}\rangle $, then by definition $\Lambda_{\chi_{\perp}}\cap \Omega=\emptyset$. It is then evident that $\Lambda_{P}\cap \Omega\neq\Lambda_{P^{\prime}}\cap \Omega$ which means $\Lambda_{P}$ and $\Lambda_{P^{\prime}}$. Hence, a non-1M$\psi$E model is  mixed-state preparation-contextual. In other words, a mixed-state preparation non-contextual model is the 1M$\psi$E one. They have also argued that converse does not hold good due to the existence of KS model \cite{leifer13}. We note here that in Leifer-Maroney proof the the  preparation non-contextuality for pure state is implicitly assumed. However, the proof does not rely on this assumption. Generalization of the above proof for arbitrary dimensional system can be found  in \cite{banik,leiferquanta}. Leifer and Maroney  have also argued that a 1M$\psi$E model is KS non-contextual but converse does not hold. 

We provide a direct and elegant proof of the Theorem-1.  We start by noting the fact that the normalization condition provides $\int_{\Lambda_{P^{\prime}}} \mu_{P^{\prime}}(\lambda|\frac{\mathcal{I}}{2})=1$. Assuming preparation non-contextuality for mixed state, i.e., $\mu_{P}(\lambda|\frac{\mathcal{I}}{2})=\mu_{P^{\prime}}(\lambda|\frac{\mathcal{I}}{2})$, we have 
\begin{align}
	\int_{\Lambda_{P^{\prime}}} \mu_{P}(\lambda|\frac{\mathcal{I}}{2})d\lambda=1
\end{align}
which, by using Eq. (\ref{ep1}) takes the form  
\begin{align}
\label{pp}
	\frac{1}{2}\int_{\Lambda_{P^{\prime}}} \mu(\lambda|\chi) d\lambda  + \frac{1}{2} \int_{\Lambda_{P^{\prime}}} \mu(\lambda|\chi_{\perp})d\lambda=1
\end{align}
Since $\Lambda_{P^{\prime}}=\Lambda_{\eta}\cup\Lambda_{\eta_{\perp}}$, Eq. (\ref{pp}) can be re-written as 

\begin{eqnarray}
\label{pnc1}
	&&\int_{\Lambda_{\eta}} \mu(\lambda|\chi) d\lambda + \int_{\Lambda_{\eta_{\perp}}}  \mu(\lambda|\chi) d\lambda\\
	\nonumber
	&+& \int_{\Lambda_{\eta}} \mu(\lambda|\chi_{\perp}) d\lambda + \int_{\Lambda_{\eta_{\perp}}} \mu(\lambda|\chi_{\perp})d\lambda=2
	\end{eqnarray}
As already mentioned that for a $\psi$-epistemic model model, $\int _{\Lambda_{\phi}} \mu(\lambda|\psi)  d\lambda = f(\psi,\phi) |\langle\psi|\phi\rangle|^{2}$ and for 1M$\psi$E model $f(\psi,\phi)=1$ where $|\psi\rangle$ and $|\phi\rangle$ are arbitrary. From Eq.(\ref{pnc1}), we then have 
\begin{eqnarray}
\label{pnc2}
&&\big[f(\chi,\eta)-f(\chi_{\perp},\eta)\Big]\langle\chi|\eta\rangle|^{2} +f(\chi_{\perp},\eta)\\ 
\nonumber
&+& \big[f(\chi,\eta_{\perp})-f(\chi_{\perp},\eta_{\perp})\Big]\langle\chi|\eta_{\perp}\rangle|^{2} +f(\chi_{\perp},\eta_{\perp})=2
\end{eqnarray} 

which can be satisfied if each of the degree of epistemicity functions involved in Eq. (\ref{pnc2}) equals to unity, i.e., $f(\chi,\eta)=f(\chi,\eta_{\perp})=f(\chi_{\perp},\eta)=f(\chi_{\perp},\eta_{\perp})=1$. This in turn proves that the model is 1M$\psi$E by definition. We thus provided a direct and simpler proof to demonstrate that mixed-state preparation non-contextuality implies 1M$\psi$E. The proof using qubit system suffices the purpose of this work but it can be generalized for any mixed state in arbitrary dimensional Hilbert space.

\section{2M$\psi$E $\Rightarrow$ pure-state preparation non-contextuality}
Next, we consider the second definition of maximal $\psi$-epistemicity (2M$\psi$E) in an ontological model of quantum theory. Let us recall the notion of probability of successful distinction of two quantum states in Hilbert space and corresponding to two epistemic states in ontic state space. Consider two arbitrary states $|\psi_{1}\rangle$ and $|\psi_{2}\rangle$ in a $d$-dimensional Hilbert space prepared by two preparation procedures $P$ and $P^{\prime}$ and their associated epistemic states are $\mu_{P}(\lambda|\psi_{1})$ and $\mu_{\mathcal{P^{\prime}}}(\lambda|\psi_{2})$  respectively. The explicit mentioning of preparation procedures $P$ and $P^{\prime}$ will be made clear soon. When respective ontic state spaces $\Lambda_{\psi_{1}}$ and $\Lambda_{\psi_{2}}$ corresponding to the states $|\psi_{1}\rangle$ and $|\psi_{2}\rangle$ have non-zero overlap, we have $\Lambda_{\psi_{1}}\cap\Lambda_{\psi_{2}}\neq\emptyset$. In such a case, $\int _{\Lambda_{\psi_{1}}}\mu_{\mathcal{P^{\prime}}}(\lambda|\psi_{2})d\lambda\neq 0 $ and similarly  $\int _{\Lambda_{\psi_{2}}} \mu_{P}(\lambda|\psi_{1})d\lambda\neq 0 $. 

In order to quantify the the overlap between  two epistemic states $\mu_{P}(\lambda|\psi_{1})$ and $\mu_{\mathcal{P^{\prime}}}(\lambda|\psi_{2})$, one can start from classical trace distance between two classical probability distributions. One can then define an appropriate quantity known as classical fidelity \cite{leifer14,barrett14,bran, nigg,ring},

\begin{align}
\label{lc1}
L_{C}(\mu_{\psi_{1}},\mu_{\psi_2})=\int _{\Lambda}min \{\mu_{P}(\lambda|\psi_{1}), \mu_{\mathcal{P^{\prime}}}(\lambda|\psi_{2})\} \ d\lambda
\end{align}
 By definition, $0\leq L_{C}(\mu_{\psi_{1}},\mu_{\psi_2})\leq 1$, and the value is $1(0)$ for identical ( disjoint) epistemic states respectively. This definition implies that the impossibility of discriminating non-orthogonal quantum states would be explained in a natural way if the two quantum states sometimes correspond to the same state of reality. But this explanation is expected to fail if the quantum and classical overlaps are not equal. This fact is utilized in \cite{maroney12, barrett14, leifer14, bran,ring} to show the inability of $\psi$-epistemic model in explaining the distinguishability between two quantum states belongs to the Hilbert space having dimension grater than two.

In quantum theory, the overlap between two states $|\psi_1\rangle$ and $|\psi_2\rangle$ can be defined as 
\begin{align}
\label{lq1}
	L_{Q}(\psi_{1},\psi_{2}) = 1 - D_{Q}(\psi_1,\psi_2)
\end{align}
where $D(\psi_1,\psi_2)=\sqrt{1-|\langle\psi_1 |\psi_2\rangle|^{2}}$ is the trace-distance. By definition, $0\leq L_{Q}(\psi_{1},\psi_{2})\leq 1$. If $|\psi_{1}\rangle$ and $|\psi_{2}\rangle$ are identical (orthogonal), $L_{Q}(\psi_{1},\psi_{2})$ takes the value $1(0)$. 

Now, since $0\leq L_{Q}(\psi_{1},\psi_{2})\leq 1$ and $0\leq L_{C}(\mu_{\psi_{1}},\mu_{\psi_2})\leq 1$, we may then categorize the possible cases are the following;

\textbf{(i)}  For \emph{any} two quantum states, if $L_{Q}(\psi_{1},\psi_{2})\neq 0$ and $L_{C}(\mu_{\psi_{1}},\mu_{\psi_2})=0$, then the corresponding ontological model is termed as \emph{$\psi$-ontic }. In this model,  different (even non-orthogonal) quantum states compatible with distinct ontic states. 

\textbf{(ii)} If a model is \emph{not} $\psi$-ontic then it is $\psi$-epistemic. There exists \emph{at least} two quantum states for which $L_{Q}(\psi_{1},\psi_{2})\neq 0$ and $L_{C}(\psi_{1},\psi_{2})\neq 0$ but $L_{Q}(\psi_{1},\psi_{2})\geq L_{C}(\mu_{\psi_{1}},\mu_{\psi_2})$. Such a model is called \emph{$\psi$-epistemic } in which epistemic states corresponding to two different quantum states has at least a common $\lambda$ and the quantum state is interpreted to have contained the information about real physical state of the system but not the reality itself. 

\textbf{(ii$a$)} A model is defined as 2M$\psi$E if  the special case of  equality in (ii) is satisfied, i.e.,  $L_{Q}(\psi_{1},\psi_{2})=L_{C}(\mu_{\psi_{1}},\mu_{\psi_2})$. In such a model, the overlap between  two epistemic states fully accounts the overlap between the two quantum states. From state discrimination viewpoints, in a 2M$\psi$E model the discrimination between two non-orthogonal quantum states is completely and quantitatively explained by the discrimination between the corresponding epistemic states.

Equipped with those definitions, we are now in a position to demonstrate the connection between  2M$\psi$E model and  preparation non-contextuality for pure-state. We prove the following theorem.  \\

\textbf{Theorem 2:} \textit{A 2M$\psi$E ontological model of quantum theory is  pure-state preparation non-contextual.} \\

 The proof of Theorem-2 is straightforward but implication of it is quite interesting. Let two preparation procedures  $P$ and ${P^{\prime}}$ prepare the \emph{same} pure state $|\psi\rangle$ in a $d-$dimensional Hilbert space. The associated epistemic states are $\mu_{P}(\lambda|\psi)$  and $\mu_{P^{\prime}}(\lambda|\psi)$ respectively.  Now, if $|\psi_1\rangle=|\psi_2\rangle \equiv |\psi\rangle$, from Eq.(\ref{lq1}) we have $D_{Q}(\psi_1,\psi_2)=0$ leading to $	L_{Q}(\psi_{1},\psi_{2}) = 1$. As defined in (ii.$a$), a 2M$\psi$E model thus demands 
\begin{align}
\label{proof1}
L_{C}(\mu_{P}(\lambda|\psi), \mu_{P^{\prime}}(\lambda|\psi)) = 1
\end{align}
This is possible only when $\mu_{P}(\lambda|\psi)=\mu_{\mathcal{P^{\prime}}}(\lambda|\psi)\equiv \mu(\lambda|\psi)$, i.e., preparation non-contextual for pure-state. This concludes the proof of Theorem-2. The above proof does not require the explicit use of response function and valid for any arbitrary dimension of the Hilbert space.    

%However, in general $L_{C}(\mu_{P_1}(\lambda|\psi), \mu_{P_2}(\lambda|\psi)) < 1$. Such a model is regarded as the non-maximally or simply $\psi-$epistemic model. Then, if $0< L_{C}(\mu_{\psi_{1}},\mu_{\psi_2})<1$, the model is pure state preparation contextual and $\psi-$epistemic. 

\section{Mixed state preparation non-contextuality $\centernot\iff$ pure state preparation non-contextuality}

A crucial point to note here that we have used the \emph{same} notion of  preparation non-contextuality assumption at the level of mixed-state in Theorem-1 and then at the level of pure-state in Theorem-2. In Theorem-1, we showed that mixed-state preparation non-contestuality $\Rightarrow$ 1M$\psi$E  and in Theorem-2 we proved 2M$\psi$E $\Rightarrow$ pure-state preparation non-contextuality. If notion of maximal $\psi$-epistemicity remains equivalent in both the definitions of 1M$\psi$E and 2M$\psi$E, we can write the following; mixed-state preparation non-contextuality $\Rightarrow$ pure-state preparation non-contextuality. However, this inference is not in accordance with the Theorem-3 whose statement is the following.\\

\textbf{Theorem 3:} \textit{The assumption of preparation non-contextuality in an ontological model of quantum theory does not hold both for mixed-state and pure-state simultaneously.} \\

To prove  Theorem-3, a particular example will suffice the purpose.  We specifically argue that  in any all-versus-nothing proof of preparation contextuality, there is an inherent interplay between the pure-state and mixed-state preparation non-contextuality in an ontological model. This was hitherto unnoticed which we made explicit here. To impose the assumption of preparation non-contextuality at the level of mixed state inevitably requires the \emph{preparation contextuality} for pure state and vice-versa. We may also remark here that since any KS proof can be cast into a proof of preparation contextuality, then a KS proof would also suffice our purpose.  But this requires minimum three dimensional Hilbert space to run the argument. We use the Spekken's \cite{spek05} proof of preparation contextuality for a qubit system.

Let $\{P_{t}\} $ is a  set of three preparation procedures in quantum theory (with $t=1,2,3$) preparing the maximally mixed state $\mathcal{I}/2$, so that 
\begin{align}
\label{pt}
	\frac{\mathcal{I}}{2}=\frac{1}{2}(A_t^{+}+A_t^{-})
\end{align}
where $A_{t}^{\pm}=(\mathcal {I}\pm A_{t})/2$ are the rank-1 projectors and $A_{t}$ are qubit observables. Now, consider that the same maximally mixed state $\mathcal{I}/2$ is also prepared by two more preparation procedures $P_4$ and $P_5$ are given by 
\begin{align}
\label{pnt}
	\frac{\mathcal{I}}{2}=\frac{1}{3}\sum_{t=1}^{3}A_t^{+}; \ \ \ \ \frac{\mathcal{I}}{2}=\frac{1}{3}\sum_{t=1}^{3}A_t^{-}
		\end{align} 
This requires the qubit observables to satisfy the relation $\sum_{t=1}^{3} A_{t}=0$.

 In an ontological model, assuming the preparation non-contextuality for the epistemic states corresponding to the mixed state $\mathcal{I}/2$ prepared by five preparations, one can write $\mu_{P_1}(\lambda|\frac{\mathcal{I}}{2})=\mu_{P_2}(\lambda|\frac{\mathcal{I}}{2})=\mu_{P_3}(\lambda|\frac{\mathcal{I}}{2})=\mu_{P_4}(\lambda|\frac{\mathcal{I}}{2})=\mu_{P_5}(\lambda|\frac{\mathcal{I}}{2})\equiv\nu(\lambda|\frac{\mathcal{I}}{2})$. The relevant ontic state spaces are respectively denoted as $\Lambda^{\mathcal{I}/2 }_{{P}_j}$ with $j=1,2,3,4,5$. Clearly, in a preparation non-contextual model, $\Lambda^{\mathcal{I}/2 }_{{P}_1}=\Lambda^{\mathcal{I}/2 }_{{P}_2}=....=\Lambda^{\mathcal{I}/2 }_{P_5}\equiv \Lambda^{\mathcal{I}/2 }$. Using convexity property and assuming preparation non-conextuality for the mixed state, one can write 
\begin{eqnarray}
\label{ptlambda}
	\nu(\lambda|\frac{\mathcal{I}}{2})&=&\frac{1}{2}\left(	\mu_{P_{t}}(\lambda|A_t^{+}) +	\mu_{P_{t}}(\lambda|A_t^{-})\right)\\
\label{p45p}
	&=&\frac{1}{3}\sum_{t=1}^{3}\mu_{P_4}(\lambda|A_t^{+})\\
	\label{p45m}
		&=&\frac{1}{3}\sum_{t=1}^{3}\mu_{P_5}(\lambda|A_t^{-}) 
\end{eqnarray}

where $t=1,2,3$. Each of the six qubit projectors appears twice in the preparation of $\rho=\mathcal{I}/2$ by those five different preparation procedures. For example, $A_{1}^{+}$ appears in $P_1$ and $P_4$. The assumption of preparation non-contextuality is equally applicable to each the rank-1 density matrices $A_{t}^{\pm}$  prepared by two distinct preparation procedures.  We denote the ontic space corresponding to the pure states as $\Lambda^{A_{t}^{\pm} }_{{P}_j}$.

Now, consider an arbitrary ontic state, say, $\lambda^{\ast}\in\Lambda^{\mathcal{I}/2 }$ for which $\nu(\lambda^{\ast}|\mathcal{I}/2 )>0$. Since $A_{t}^{+}$ and $A_{t}^{-}$ are orthogonal then $\Lambda_{P_t}^{A_{t}^{+} }\cap \Lambda_{P_t}^{A_{t}^{-} } =\emptyset$. If we consider $\mu_{P_{t}}(\lambda^{\ast}|{A_t}^{+})>0$ for every $t$ then   $\mu_{P_{t}}(\lambda^{\ast}|{A_t}^{-})=0$. If preparation non-contextuality for every pure state is also assumed (for example, $\mu_{P_{t}}(\lambda^{\ast}|A_t^{+})=\mu_{P_{4}}(\lambda^{\ast}|A_t^{+}) )$, then from Eq. (\ref{p45m}) one has $\nu(\lambda^{\ast}|\mathcal{I}/2 )=0$.  Since $\nu(\lambda^{\ast}|\mathcal{I}/2 )>0$,  we thus have a contradiction. The argument holds for any $\lambda$ and thus similar contradiction can be found for any such assignment of positive probability distribution. 

By the word contradiction, we  explicitly mean here that if one assumes preparation non-contextuality for pure states then the epistemic states corresponding to the mixed states become preparation contextual. On the other hand, if one wishes to assume preparation contextuality for pure states (for example, given a $\lambda^{\ast}$  the epistemic states corresponding to the pure states $\mu_{P_{t}}(\lambda^{\ast}|\{\rho_{A_t}^{\pm}\})$ change their support in the preparation procedures $P_{5}$ or $P_{4}$) then the preparation non-contextuality for mixed state can be maintained. Thus, the assumption of preparation noncontextuality cannot be imposed for both pure-state and mixed state together, which is the statement of Theorem-3. However, there is no restriction for both pure and mixed-state to be preparation contextual in an ontological model of quantum theory. 

\section{Summary and discussion}
In summary, we provided three theorems regarding the connections between the notions of preparation non-contextuality and maximal $\psi$-epistemicity in an ontological model. We showed that,  taken together, those three theorems  leads conceptually inconsistent conclusions. Note that the notion preparation non-contextuality in an ontological model ($\mu_{P}(\lambda|\rho)=\mu_{P^{\prime}}(\lambda|\rho))$ usually  corresponds to a mixed state $\rho$ prepared by two or more distinct procedures $P$ and $P^{\prime}$.  Although it not hitherto spelled out explicitly, but there is  whatsoever no reason to constrain the notion of preparation non-contextuality to not to be applicable for pure-state $\rho=|\psi\rangle\langle\psi|$ prepared by distinct procedures.  As mentioned earlier, the maximal $\psi$-epistemicity in an ontological model of quantum theory conceptually implies that  the overlap between any two quantum states in Hilbert space is fully accounted for by the overlap of their respective epistemic states in ontic state space. Interestingly, the same notion of maximal $\psi$-epistemicity is captured by two distinct mathematical definitions, termed here as 1M$\psi$E and 2M$\psi$E. In the very first paper on maximal $\psi$-epistemicity by Maroney \cite{maroney12}, 2M$\psi$E was introduced as a noise tolerant version of 1M$\psi$E. 

In Theorem-1, using the first mathematical definition of maximal $\psi$-epistemicity we showed that mixed-state preparation non-contextuality $\Rightarrow$ 1M$\psi$E. This theorem was proved in \cite{leifer13} but we provided here a simpler and direct proof of it. By using the second  mathmatical definition of maximal $\psi$-epistemicity we proved the Theorem-2 to demonstrate that 2M$\psi$E $\Rightarrow$ pure-state preparation non-contextuality.  Now, if notion of maximal $\psi$-epistemicity is equivalent in both the definitions of 1M$\psi$E and 2M$\psi$E, one can infer that the mixed-state preparation non-contextuality $\Rightarrow$ pure-state preparation non-contextuality.  However,  we argued in Theorem-3 that there are scenarios (for example, any all-versus-nothing proof of preparation contextuality) where the assumption of preparation non-contextuality cannot be imposed for the epistemic states corresponding to both the pure and mixed states. To maintain the preparation non-contextuality for a mixed state one has to allow the preparation contextuality for pure state and vice versa. Thus, the statement of the Theorem-3 is in clear disagreement with the conclusion drawn from the first two theorems. One may thus conclude that the notion of maximal $\psi$-epistemicity captured by 1M$\psi$E and 2M$\psi$E are inequivalent and 2M$\psi$E is merely not a noise tolerant version of 1M$\psi$E.  We may note here that, it will cerrtainly be interesting if a direct proof of the inequivalence between  1M$\psi$E and 2M$\psi$E can be demonstrated.

Finally, we make a few remarks about further implications of our work. Note that,  a mixed-state preparation non-contextual model and 1M$\psi$E model both are outcome deterministic \cite{spekkens13, maroney12, bal}.  Also, the deterministic response function is a key element for connecting 1M$\psi$E and KS non-contextuality which eventually proves mixed-state preparation non-contextuality implies KS non-contextuality \cite{leifer13}. On the other hand, 2M$\psi$E modes does not use the response function and it is remained to be examined whether 2M$\psi$E model is deterministic or not. However, a pure-state preparation non-contextual model is not expected to be a deterministic theory as can be understood from Beltrametti-Bugajski model \cite{bel}. This indicates that there is no obvious connection between 2M$\psi$E and KS non-contextuality. 

The 2M$\psi$E model can provide an insight into the no-go theorem proposed by Hardy \cite{hardy}. He introduced a $\psi$-ontology theorem by considering a key assumption known as `ontic indifference'. The assumption of ontic indifference states the following. If under quantum transformation a pure state $|\psi\rangle$ remains unaffected  (i.e., $\mathcal{U}|\psi\rangle=|\psi\rangle$)  then suitable transformation in the ontic space can be found so that every ontic state in the support of epistemic state corresponding to $|\psi\rangle$ remain unaffected by the transformation.  Now, by considering $\mathcal{U}|\psi\rangle=|\psi\rangle$ is an another preparation procedure to prepare $|\psi\rangle$, a weaker reading of ontic indifference assumption could be the pure-state preparation non-contextuality where epistemic state is assumed to remain unaffected by the transformation. This indicates an interesting connection between 2M$\psi$E model and Hardy's ontological model satisfying a weaker ontic indifference assumption, which calls for further study. 

It could also be an interesting study to examine the limit on the degree of preparation contextuality (for pure or mixed state) that can be imposed by quantum theory. For example, in the Bell-CHSH scenario, mixed-state preparation non-contextuality implies the locality condition \cite{pusey,tava}. But, quantum violation is restricted to Cirelson bound and hence put a limit on the degree of mixed-state preparation contextuality. In Bell-CHSH scenario, there is no issue of pure-state preparation contextuality because same pure state does not appear in different context. One may quantify the degree of preparation contextuality for pure and mixed state in suitable scenario. This requires to introduce some new ingredients where such a degree is meaningful and can provides some interesting conclusions. Such a degree of preparation contextuality can also be compared with the degree of $\psi$-epistemicity. This may provide new insights into the research on  $\psi$-epistemic models. Further studies along this line could be an interesting avenue for future research. 
\section*{Acknowledgments}
 Author acknowledges the support from the project DST/ICPS/QuEST/2018/Q-42.

\appendix{}

\end{document}